# Effects Investigation Of MAC And PHY Layer Parameters On The Performance Of IEEE 802.15.6 CSMA/CA

Md. Abubakar Siddik, Most. Anju Ara Hasi, Md. Rajiul Islam and Jakia AkterNitu

Department of Electronics and Communication Engineering, Hajee Mohammad Danesh Science and Technology University, Dinajpur, Bangladesh

*ABSTRACT*

*The recently released IEEE 802.15.6 standard specifies several physical (PHY) layer and medium access control (MAC) layer protocols for variety of medical and non-medical applications of Wireless Body Area Networks (WBAN). The most suitable way for enhancing network performance is to be the choice of different MAC and PHY parameters based on quality of service (QoS) requirements of different applications. The impact of different MAC and PHY parameters on the network performance and the trade-off relationship between the parameters are essential to overcome the limitations of exiting carrier sense multiple access with collision avoidance (CSMA/CA) scheme of IEEE 802.15.6 standard. To address this issue, we develop a Markov chain-based analytical model of IEEE 802.15.6 CSMA/CA for all user priorities (UPs) and apply this general model to different network scenarios to investigate the effects of the packet arrival rate, channel condition, payload size, access phase length, access mechanism and number of nodes on the performance parameters viz. reliability, normalized throughput, energy consumption and average access delay. Moreover, we conclude the effectiveness of different access phases, access mechanisms and user priorities of intra-WBAN.*

## 1. INTRODUCTION

In the era of advanced technology, increasing needs of aging population, rising costs of healthcare, limited healthcare resources, especially during worldwide pandemic like COVID-19, have triggered the concepts of Wireless Body Area Networks (WBANs) as a primary part of the ubiquitous Internet of Medical Things (IoMT) systems and received considerable attention in the academy and industry. The WBAN is composed of a limited number of tiny, low-power, low-cost, wearable or implantable, intelligent, and heterogeneous medical sensors that are deployed in, on or around the proximity of the human body for continuously sensing vital physiological signals. The sensed signals are then aggregated at a coordinator via a short-range, low power wireless communication, provided by IEEE 802.15.6 standard [1] and forwarded to the servers for further analysis. It offers numerous medical and non-medical applications in ubiquitous healthcare, military and defence, sports and fitness, and entertainment fields for improving the quality of human life, described in detail in [2] – [4]. Each WBAN application has some specific QoS requirements like reliability, latency, security, and power consumption [5]. The communication architecture of WBANs is the combination of three different tiers: Tire-1, known as intra-WBAN communication, Tire-2, known as inter-WBAN communication, and Tire-3, and known as beyond-WBAN communication [5] – [7]. The IEEE 802.15 Task Group 6 has developed a new communication standard, IEEE 802.15.6 that specifies the PHY and MAC layers specifications for WBAN. The IEEE 802.15.6 standard defines MAC sub-layer that acts as a lower sub-layer of the data link layer of the OSI model. It utilizes two types of access mechanisms including: contention-based access and contention-free access. The contention-based CSMA/CA mechanism used in IEEE 802.15.6 standard has significant differences with the CSMA/CA mechanism of the others wireless communication standards viz. IEEE 802.11, IEEE 802.11e and IEEE 802.15.4 [7] – [9]. These differences make the necessity of a new analytical model to evaluate the performance of the IEEE 802.15.6 CSMA/CA. The overall performance of the WBAN mainly depends on the parameters of MAC and PHY layer, defined by IEEE 802.15.6 standard. To fulfil the desired QoS requirements and maintaining balance among the trade-off performance metrics for specific applications of WBAN by selecting optimal parameters has still the major issue. In the IEEE 802.15.6 CSMA/CA mechanism, the major parameters of MAC and PHY layers that could affect its performance include: packet arrival rate, channel condition, payload size, access phase length, number of nodes, access mechanism, traffic conditions and traffic differentiation. To the best of our knowledge, none of the current research works have considered all the above-mentioned parameters in modelling the IEEE 802.15.6 CSMA/CA and investigated the effects of these parameters on the performance metrics, viz. reliability, normalized throughput, energy consumption, and average access delay together.

In this paper, we present an extensive and comprehensive evaluation of IEEE 802.15.6 CSMA/CA through analytical analysis using Markov chain. Moreover, we identify the key parameters of MAC and PHY layer that have significant impact on overall system performance. The main contributions of this paper are summarized as

- We present a comprehensive view of the comparisons that describes the limitations, assumptions and findings of existing Markov chain-based analytical models of IEEE 802.15.6 CSMA/CA MAC protocol in WBAN.
- We develop an analytical model using 2-D Markov chain for IEEE 802.15.6 CSMA/CA MAC protocol and derive the relationship among channel idle probability, channel access probability, transmission probability, successful transmission probability, collision transmission probability, error transmission probability and failure transmission probability.
- We derive the expressions of all performance metrics viz. reliability, normalized throughput, energy consumption and average access delay to find more accurate value.
- We solve the derived complex analytical model using Maple.
- We investigate the effects of packet arrival rate, channel condition, payload size, access phase length, access mechanism and number of nodes on performance metrics.
- We examine the importance of all UPs and all APs. In addition, we investigate the effectiveness of RTS/CTS access mechanism over basic access mechanism.

The rest of the paper is organized as follows: Section 2 addresses the related works and Section 3 brief overviews the IEEE 802.15.6 standard. In Section 4, a Markov chain-based analytical model for IEEE 802.15.6 CSMA/CA is described. The expressions for different performance metrics are derived in Section 6. The performance metrics evaluation of this work is presented in Section 7. Finally, Section 8 concludes the paper and gives the future work outlines.

## 2. RELATED WORKS

There are several literatures [9] – [20] that explore the performance analysis of IEEE 802.15.6 CSMA/CA using Markov chain model. In [9], the authors proposed an analytical model for all UPs to investigate the impact of access phase length on throughput and average backoff time in saturated traffic condition. The effects of number of nodes and packet arrival rate are not mentioned in their evaluation part and two important performance metrics, reliability and energy consumption are not considered. In [10], Sarkar et al. proposed a user priority wise and Markov chain-based analytical model of IEEE 802.15.6 CSMA/CA for noisy channel and situation traffic condition to evaluate some performance parameters viz. reliability, throughput, energy consumption and average delay by taking into account the acknowledgement time or timeout after transmitting packets. The effects of channel condition, packet size and data rate on the performance metrics are also presented. In [11], user priority wise DTMC analysis was performed to evaluate the performance of slotted CSMA/CA in IEEE 802.15.6 standard for noisy channel and situation traffic condition. However, slotted CSMA/CA mechanism is not specified in IEEE 802.15.6 standard. Quan et al. [12] developed a Markov chain-based statistical model to analyse the performance of IEEE 802.15.6 CSMA/CA for noisy channel environment and non-saturation traffic condition whereas Markov Chain Monte Carlo (MCMC) method used to determine the access probabilities of nodes. The error correction capability of different code rates is mainly investigated through measuring throughput and access probability of the nodes. In [13], a Markov chain-based mathematical model was designed to evaluate the throughput, energy consumption and energy efficiency of IEEE 802.15.6 CSMA/CA for ideal channel environment and non-saturation traffic condition. T. Suzuki [14] introduced equilibrium point analysis (EPA) technique to calculate throughput and average response time of the IEEE 802.15.6 CSMA/CA from a Markov chain-based analytical model which was designed for ideal channel and non-saturation traffic condition. However, the authors [12], [14] have not considered user priority feature to evaluate the performance of CSMA/CA. As per the IEEE 802.15.6 standard, the random-access mechanism is used in different contention-based access phases (EAPs, RAPs and CAP) under beacon mode with superframes. However, in [10] – [14], [20] – [22], the authors evaluated the performance of CSMA/CA without considering APs duration and superframe structure. Jacob et al. [15] developed a sleep mechanism to extend the network lifetime and designed a user priority wise and Markov chain-based analytical model to measure the performance of IEEE 802.15.6 CSMA/CA for ideal channel environment and non-saturation traffic condition. Khan et al. [16] evaluated normalized throughput and mean frame service time of CSMA/CA mechanism by developing a user priority wise and Markov chain-based analytical model and assuming that the channel is ideal and nodes contain non-saturated traffic. Yuan et al. [17] analysed the performance of IEEE 802.15.6-based WBAN in presence of intra- and inter-WBAN interference by designing a user priority wise and Markov chain-based analytical model for ideal channel and saturation traffic condition of nodes. The above studies [10] – [12], [14] – [17], have not considered hidden and exposed terminals that create inter-BAN interference. Use of basic access mechanism of CSMA/CA with the presence of hidden and exposed terminals cause of more collisions and consequence the degradation the performance of IEEE 802.15.6 CSMA/CA. The RTS/CTS access mechanism can mitigate the impact of hidden and exposed terminals on CSMA/CA mechanism by reducing collision [24]. As discussed in [23], [25] the RTS/CTS access mechanism has potentiality to enhance the performance of the IEEE 802.15.6 CSMA/CA instead of basic access mechanism. In [18], [19] the authors designed a Markov Chain-based complete analytical model of IEEE 802.15.6 CSMA/CA by considering most of the constrains of IEEE 802.15.6 standard and evaluated the performance parameters viz. normalized throughput, successful transmission probability, mean waiting time and average time between two successive access.

## 3. IEEE 802.15.6 STANDARD

The IEEE 802.15.6 standard defines three physical (PHY) layers, i.e., narrowband (NB), ultra-wideband (UWB), and human body communications (HBC) layers. Based on the traffic designation, the IEEE 802.15.6 standard introduces eight user priorities ($UP_i$ where $i \epsilon [0,7]$) which are differentiated by $CW_{i,min}$ and $CW_{i,max}$. Traffic designation of each $UP_i$ and the relation between Contention Window bounds and $UP_i$ are depicted in Table 1. It defines three access modes: beacon mode with superframes, non-beacon mode with superframes, and non-beacon mode without superframes. The superframe of beacon mode is divided into seven access phases, i.e., EAP1, RAP1, MAP1, EAP2, RAP2, MAP2, and CAP, shown in Figure 1. The EAP1 and EAP2 are used for $UP_7$, and the RAP1, RAP2 and CAP are used for all UPs. In order to access the channel, this standard also defines three access mechanisms: random access mechanism, improvised and unscheduled access mechanism, and scheduled access mechanism. According to CSMA/CA mechanism of IEEE 802.15.6 standard, a $UP_i$ node shall maintain a backoff counter and contention window ($W_{i,j}$) to get access of the channel. The node shall set its backoff counter value to a randomly chosen integer over $[1, W_{i,j}]$ and decrease its backoff counter by one for each

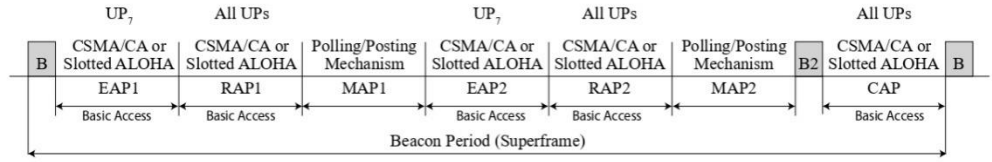

Figure 1. Layout of access phases, permitted UPs and access mechanisms for beacon mode

Table 1. MAC parameters and traffic designation of Ups

| $UP_i$ | $CW_{i,min}$ | $CW_{i,max}$ | $m_i$ | $x_i$ | Traffic designation |
|---|---|---|---|---|---|
| 0 | 16 | 64 | 4 | 3 | Background (BK) |
| 1 | 16 | 32 | 2 | 5 | Best effort (BE) |
| 2 | 8 | 32 | 4 | 3 | excellent effort (EE) |
| 3 | 8 | 16 | 2 | 5 | Control load (CL) |
| 4 | 4 | 16 | 4 | 3 | Video (VI) |
| 5 | 4 | 8 | 2 | 5 | Voice (VO) |
| 6 | 2 | 8 | 4 | 3 | High priority medical data or network control |
| 7 | 1 | 4 | 2 | 5 | Emergency or medical important report |

idle CSMA slot. The node will transmit one frame over the channel if backoff counter reaches zero. The contention window selection, locking and unlocking process of the backoff counter during the contention period depends on channel and transmission state. A new contention window is selected as

- $UP_i$ shall set $W_{i,j}$ to $CW_{i,min}$ for frame successful transmission or a new frame.
- If $UP_i$ failed to transmit, the $W_{i,j}$ unchanged if $j$ is an odd; otherwise, $W_{i,j}$ is doubled.
- $UP_i$ shall set $W_{i,j}$ to $CW_{i,max}$ if CW exceeds $CW_{i,max}$ and attempt do not exceed retry limits.

When any of the following events occurs, the node shall lock its backoff counter.
- The channel is sensed busy due to another transmissions.
- The current access phase does not permit the $UP_i$ node for the current transmission attempt.
- The current time is in the permitted access phase but the time between the end of the CSMA slot and the end of the permitted access phase is not long enough for a frame transmission.

The backoff counter will be unlocked when the channel has been idle for SIFS time within the permitted access phase and the time interval between the current time and the end of the permitted access phase is long enough for a frame transmission. After unlocking the backoff counter, the node shall start the backoff procedure and transmit frame if backoff counter reaches zero. The backoff procedure of IEEE 802.15.6 CSMA/CA mechanism is given in Figure 2.

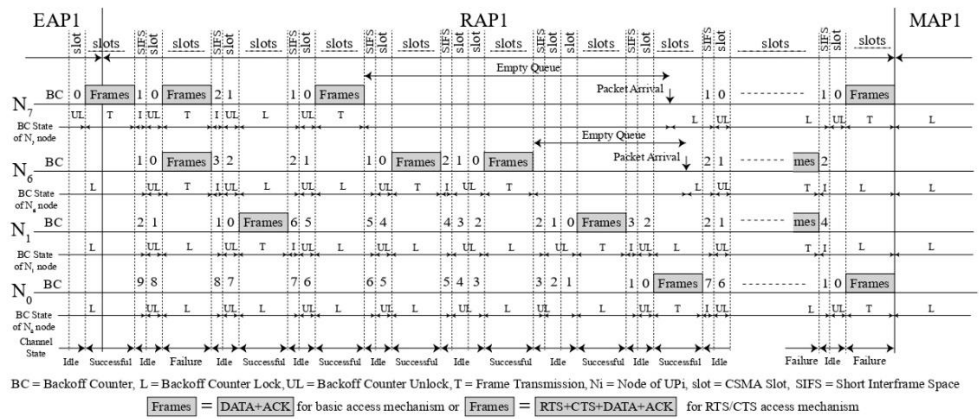

Figure 2. Backoff procedure of IEEE 802.15.6 CSMA/CA mechanism

## 4. ANALYTICAL MODEL

### 4.1. System Model

In this paper, we consider a continuous healthcare monitoring system where a patient is equipped with one and only one central hub as coordinator and up to $n$ ($n = mMaxBANSize = 64$) number of identical medical sensor nodes, deployed on the body, which together forms a one-hop star topology of intra-WBAN, shown in Figure 3 . We mainly focus on the uplink frame transmission (from node to coordinator). The network consists of all eight UPs traffic ranging from 0 to 7, where 0 denotes the lowest priority and 7 denotes highest priority traffic. It is also considered that each node has a single queue that contains only one user priority data frame. The packets arrival process of all UPs is Poisson process with rate λ and this arrival rate is equal for all UPs. We assume that the coordinator operates in beacon mode with superframe boundaries where MAP1, MAP2, EAP2, RAP2 and CAP are set to zero. It is considered that all nodes and coordinator follow immediate acknowledgement (I-ACK) policy and use automatic repeat request (ARQ) as an error control method. A failure transmission occurs due to two reasons. The first reason is collision transmission that happens when more than one node transmits at the same time and the second one is error transmission that occurs due to a noisy channel. A frame is dropped when the number of failure transmission exceeds the finite retransmission limits ($m_i + x_i = 7$). A node that has a data frame to transmit cannot generate a new data frame until either it receives the ACK frame for the data frame or the data frame is dropped. We also assume that all UPs nodes have equal traffic load and payload size and the collision probability of a frame that transmitted by a node is independent of the number of retransmissions, i.e., backoff stages. We assume that a node transmits just one data frame after successfully access the channel. We use narrowband (NB) as a PHY layer to evaluate the performance of IEEE 802.15.6 CSMA/CA. In this work, we ignore the hidden terminal, expose terminal and channel capture effect.

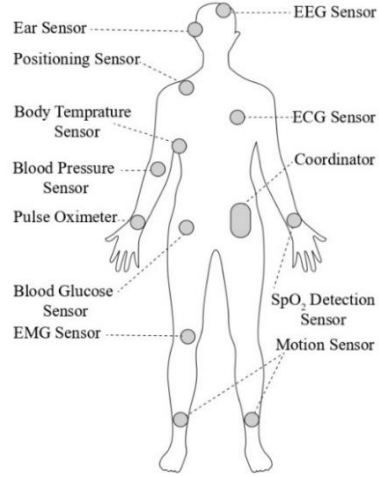

Figure 3. System model

### 4.2. Markov Chain Model

We first design a 2-D Markov chain model to describe the backoff procedure of IEEE 802.15.6 CSMA/CA for both saturation and non-saturation traffic conditions according to $M/G/1$ queuing model, which is shown in Figure 4. There are some studies where Markov chain-based analytical model was presented to evaluate the performance of CSMA/CA mechanism [9] – [12], [15] – [19], [26] – [28], [31-35]. In this Markov chain, the state of each node of $UP_i$ is represented by $(i, j, k)$ where $i, j$ and $k$ denotes the user priority of node, number of backoff stage and backoff counter value, respectively. The initial value of $j$ for a new frame is zero and is incremented by one after every failure transmission until it reaches the retry limit $(m_i + x_i)$. After every successful transmission or frame drop, the value of $j$ will be reset to zero. The value of $k$ is initially set with a value that is randomly chosen from $[1, W_{i,j}]$, where $W_{i,j}$ denotes the contention window size of $UP_i$ node at backoff stage $j$. The backoff counter value $k$ is decremented by one if the channel is sensed idle in a CSMA slot and if there is sufficient time to complete current frame transmission before the end of the current access phase. When the counter value becomes zero, frame is transmitted immediately. If failure transmission is occurred due to collision transmission or error transmission, the node goes to the next backoff stage with a new backoff counter value. If frame is successfully transmitted, the node selects a new backoff counter value under initial backoff stage if it has at least one packet for transmission. The contention window size of $UP_i$ node at backoff stage $j$ is given by

$$W_{i,j} = \begin{cases} CW_{i,min} & ; \quad j = 0 \\ 2^{j/2} \times CW_{i,min} & ; \quad 1 \leq j \leq m_i \text{ and } j \text{ is even} \\ 2^{(j-1)/2} \times CW_{i,min} & ; \quad 1 \leq j \leq m_i \text{ and } j \text{ is odd} \\ CW_{i,max} & ; \quad m_i < j \leq m_i + x_i \end{cases} \quad (1)$$

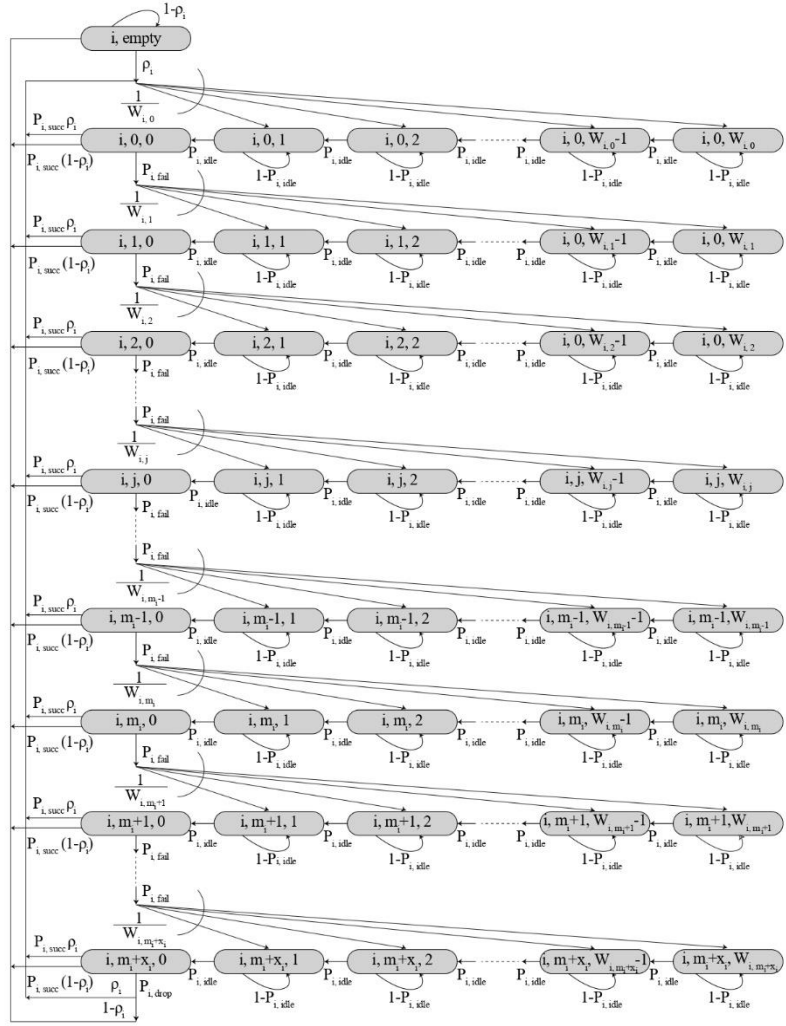

Figure 4. Markov chain for IEEE 802.15.6 CSMA/CA of $UP_i$ node

To analyse non-saturation traffic conditions of $UP_i$ node, we introduce a state in the Markov chain is denoted by $(i, empty)$, which represents the state of $UP_i$ node when the node is empty after a successful transmission or a frame drop. We define $b_{i,j,k}$ and $b_{i,empty}$ that represent the stationary probabilities of states $(i, j, k)$ and $(i, empty)$, respectively. We also define $\rho_i$ as the probability that the queue of $UP_i$ node has at least one frame waiting for transmission. The one-step transition probabilities between the states of the Markov chain are represented as

$$\begin{cases} P_{(i,empty)|(i,j,k)} = (1 - \rho_i)P_{i,succ} & ; \quad 0 \leq j \leq m_i + x_i - 1 \\ P_{(i,empty)|(i,m_i+x_i,0)} = 1 - \rho_i \\ P_{(i,empty)|(i,empty)} = 1 - \rho_i \\ P_{(i,0,k)|(i,empty)} = \dfrac{\rho_i}{W_{i,0}} & ; \quad 1 \leq k \leq W_{i,0} \\ P_{(i,0,k)|(i,j,k)} = \dfrac{\rho_i}{W_{i,0}} P_{i,succ} & ; \quad 0 \leq j \leq m_i + x_i - 1 \text{ and } 1 \leq k \leq W_{i,0} \\ P_{(i,0,k)|(i,j,0)} = \dfrac{\rho_i}{W_{i,0}} & ; \quad 1 \leq k \leq W_{i,0} \end{cases} \quad (2)$$

$$P_{(i,j,k)|(i,m_i+x_i,0)} = 1 - P_{i,idle} \quad ; \quad 0 \leq j \leq m_i + x_i - 1 \text{ and } 1 \leq k \leq W_{i,j}$$

$$P_{(i,j,k)|(i,j-1,k)} = P_{i,idle} \quad ; \quad 1 \leq k \leq W_{i,j}$$

$$P_{(i,j,k)|(i,j,k-1)} = \frac{P_{i,fail}}{W_{i,j}} \quad ; \quad 0 \leq j \leq m_i + x_i - 1 \text{ and } 1 \leq k \leq W_{i,j}$$

We assume that the retry limit of CSMA/CA mechanism is finite. When number of failure transmission of $UP_i$ node exceed the retry limit $(m_i + x_i)$, the frame is dropped and the node initiates the backoff procedure again to transmit a new frame. Thus, the frame drop probability is given as

$$P_{i,drop} = (P_{i,fail})^{m_i+x_i+1} \tag{3}$$

Now, we derive the stationary probabilities $b_{i,j,k}$ and $b_{i,empty}$ in terms of $b_{i,0,0}$ state. According to the Markov chain, the stationary probability of each state for the zero backoff stage can be written as

$$\begin{cases}
b_{i,0,W_{i,0}} = (1 - P_{i,idle})b_{i,0,W_{i,0}} + \sum_{j=0}^{m_i+x_i-1} \frac{\rho_i P_{i,succ}}{W_{i,0}} b_{i,j,0} + \frac{\rho_i}{W_{i,0}} b_{i,m_i+x_i,0} + \frac{\rho_i}{W_{i,0}} b_{i,empty} \\
b_{i,0,W_{i,0}-1} = P_{i,idle}b_{i,0,W_{i,0}} + (1 - P_{i,idle})b_{i,0,W_{i,0}} + \sum_{j=0}^{m_i+x_i-1} \frac{\rho_i P_{i,succ}}{W_{i,0}} b_{i,j,0} \\
\quad\quad + \frac{\rho_i}{W_{i,0}} b_{i,m_i+x_i,0} + \frac{\rho_i}{W_{i,0}} b_{i,empty} \\
\quad\quad\quad\quad\quad\quad\quad\quad\quad\quad\quad\quad\vdots \\
b_{i,0,2} = P_{i,idle}b_{i,0,3} + (1 - P_{i,idle})b_{i,0,2} + \sum_{j=0}^{m_i+x_i-1} \frac{\rho_i P_{i,succ}}{W_{i,0}} b_{i,j,0} + \frac{\rho_i}{W_{i,0}} b_{i,m_i+x_i,0} \\
\quad\quad + \frac{\rho_i}{W_{i,0}} b_{i,empty} \\
b_{i,0,1} = P_{i,idle}b_{i,0,2} + (1 - P_{i,idle})b_{i,0,1} + \sum_{j=0}^{m_i+x_i-1} \frac{\rho_i P_{i,succ}}{W_{i,0}} b_{i,j,0} + \frac{\rho_i}{W_{i,0}} b_{i,m_i+x_i,0} \\
\quad\quad + \frac{\rho_i}{W_{i,0}} b_{i,empty}
\end{cases} \tag{4}$$

By solving all the equations in Eq. (4), the general expression of the stationary probabilities for zero backoff stage can be written in terms of $b_{i,0,0}$ state, i.e.,

$$b_{i,0,k} = \frac{W_{i,0} - k + 1}{W_{i,0}} \frac{1}{P_{i,idle}} b_{i,0,0} \quad ; \quad \mathbf{1 \leq k \leq W_{i,0}} \tag{5}$$

Again, from the Markov chain, the stationary probability of each state for the $j^{th}$ backoff stage $(1 \leq j \leq m_i + x_i)$ can be written as

$$\begin{cases}
b_{i,j,W_{i,j}} = \frac{P_{i,fail}}{W_{i,j}} b_{i,j-1,0} + (1 - P_{i,idle})b_{i,j,W_{i,j}} \\
b_{i,j,W_{i,j}-1} = P_{i,idle}b_{i,j,W_{i,j}} + \frac{P_{i,fail}}{W_{i,j}} b_{i,j-1,0} + (1 - P_{i,idle})b_{i,j,W_{i,j}-1} \\
\quad\quad\quad\quad\quad\quad\quad\quad\quad\quad\vdots \\
b_{i,j,2} = P_{i,idle}b_{i,j,3} + \frac{P_{i,fail}}{W_{i,j}} b_{i,j-1,0} + (1 - P_{i,idle})b_{i,j,2} \\
b_{i,j,1} = P_{i,idle}b_{i,j,2} + \frac{P_{i,fail}}{W_{i,j}} b_{i,j-1,0} + (1 - P_{i,idle})b_{i,j,1}
\end{cases} \tag{6}$$

By solving all the equations in Eq. (6), the general expression of stationary probabilities for $j^{th}$ backoff stage can be written in terms of $b_{i,0,0}$ state, i.e.,

$$b_{i,j,k} = \frac{W_{i,j} - k + 1}{W_{i,j}} \frac{P_{i,fail}}{P_{i,idle}} b_{i,j-1,0} \qquad ; \quad \mathbf{1 \leq j \leq m_i + x_i} \text{ and } \mathbf{1 \leq k \leq W_{i,j}} \qquad (7)$$

According to the one-step transition probability of the Markov chain in Eq. (2), we have

$$b_{i,j,0} = P_{i,idle} b_{i,j,1} \qquad ; \quad 1 \leq j \leq m_i + x_i \qquad (8)$$

By adding all the equations in Eq. (6) and Eq. (8), we have

$$b_{i,j,0} = P_{i,fail} b_{i,j-1,0} \qquad ; \quad 1 \leq j \leq m_i + x_i \qquad (9)$$

Now we explore the Eq. (9) for the all values of $j$ ($1 \leq j \leq m_i + x_i$) and we have

$$\begin{cases} b_{i,1,0} = P_{i,fail} b_{i,0,0} \\ b_{i,2,0} = P_{i,fail} b_{i,1,0} \\ \vdots \\ b_{i,m_i+x_i-1,0} = P_{i,fail} b_{i,m_i+x_i-2,0} \\ b_{i,m_i+x_i,0} = P_{i,fail} b_{i,m_i+x_i-1,0} \end{cases} \qquad (10)$$

By solving all the equations in Eq. (10), the last state of each backoff stage can be also written in terms of $b_{i,0,0}$ state, i.e.,

$$b_{i,j,0} = (P_{i,fail})^j b_{i,0,0} \qquad ; \quad 1 \leq j \leq m_i + x_i \qquad (11)$$

By substituting Eq. (5) and Eq. (10) in Eq. (7), the general expression of the stationary probability for each backoff stage of the Markov chain can be obtained in terms of $b_{i,0,0}$ state, i.e.,

$$b_{i,j,k} = \frac{W_{i,j} - k + 1}{W_{i,j}} \frac{(P_{i,fail})^j}{P_{i,idle}} b_{i,0,0} \qquad ; \quad \mathbf{0 \leq j \leq m_i + x_i} \text{ and } \mathbf{1 \leq k \leq W_{i,j}} \qquad (12)$$

The stationary probability of empty state can be obtained from the Markov chain as

$$b_{i,empty} = (1 - \rho_i) b_{i,empty} + \sum_{j=0}^{m_i+x_i-1} (1 - \rho_i) P_{i,succ} b_{i,j,0} + (1 - \rho_i) b_{i,m_i+x_i,0} \qquad (13)$$

By substituting Eq. (11) in Eq. (12), the stationary probability of empty state can be obtained in terms of $b_{i,0,0}$ state, i.e.,

$$b_{i,empty} = \frac{(1 - \rho_i)}{\rho_i} b_{i,0,0} \qquad (14)$$

Now, the normalization condition of the Markov chain can be expressed as

$$\sum_{j=0}^{m_i+x_i} b_{i,j,0} + \sum_{k=1}^{W_{i,j}} b_{i,0,k} + \sum_{j=1}^{m_i+x_i} \sum_{k=1}^{W_{i,j}} b_{i,j,k} + b_{i,empty} = 1 \qquad (15)$$

By substituting Eq. (11), Eq. (12) and Eq. (14) in Eq. (15), the stationary probability of $(i,0,0)$ state can be obtained, i.e.,

$$b_{i,0,0} = \frac{1}{\frac{(1-P_{i,fail})^{m_i+x_i+1}}{1-P_{i,fail}} + \sum_{j=0}^{m_i+x_i} \frac{W_{i,j}+1}{2} \frac{(P_{i,fail})^j}{P_{i,idle}} + \frac{(1-\rho_i)}{\rho_i}} \qquad (16)$$

According to the CSMA/CA mechanism, a node will attempt for transmission when the node is in $(i,j,0)$, $0 \leq j \leq m_i + x_i$, state of the Markov chain. Thus, the transmission probability of each $UP_i$ node is written as

$$\tau_i = \sum_{j=0}^{m_i+x_i} b_{i,j,0} \qquad (17)$$

By substituting Eq. (11) in Eq. (17), the transmission probability can be obtained in terms of $b_{i,0,0}$ state, i.e.,

$$\tau_i = \frac{(1-P_{i,fail})^{m_i+x_i+1}}{1-P_{i,fail}} b_{i,0,0} \qquad (18)$$

International Journal of Wireless & Mobile Networks (IJWMN), Vol.15, No.6, December 2023According to the CSMA/CA mechanism of IEEE 802.15.6 standard, we define $P_{i,lock}$ as the probability that in a given CSMA slot if there is not enough time for transmitting a frame in the current permitted access period. At this time the backoff counter shall be freeze till the beginning of the next permitted access period. Thus, the counter lock probability is estimated as

$$P_{i,lock} = \begin{cases} \dfrac{1}{L_{rap} - L_{succ} - C_i} & ; \quad i = 7 \\ \dfrac{1}{L_{eap} + L_{rap} - L_{succ} - C_i} & ; \quad 0 \leq i \leq 6 \end{cases} \quad (19)$$

where $L_{eap}$, $L_{rap}$ and $L_{succ}$ denote the length of EAP in slots, the length of RAP in slots and the duration of successful transmission time in slots, respectively. The mean backoff value of $UP_i$ node ($C_i$) is approximated as

$$C_i = \frac{\frac{m_i}{2} + W_{i,j}\left(2^{\frac{m_i}{2}} - 1\right) + (x_i + 1)\frac{W_{i,j}2^{\frac{m_i}{2}+1}}{2}}{m_i + x_i + 1} \quad ; \quad 0 \leq i \leq 7 \quad (20)$$

The transmission probability in a CSMA slot is that at least one node transmits a frame through the channel. Thus, the transmission probability in EAP1 and in RAP1 is denoted by $P_{tran,eap}$ and $P_{tran,rap}$, respectively and that are determined as

$$\begin{cases} P_{tran,eap} = 1 - (1 - \tau_7)^{n_7} \\ P_{tran,rap} = 1 - \displaystyle\prod_{i=0}^{7}(1 - \tau_i)^{n_i} \end{cases} \quad (21)$$

Where $n_i$ denotes the number of nodes of $UP_i$ in the network. We assume that number of nodes of each user priority is equal, $n_i = n/n_{UP}$, where $n$ is the total number of nodes in the network and $n_{UP}$ is the number of user priorities. The idle probability of a CSMA slot is that no node is transmitting in the slot. Thus, we have

$$\begin{cases} P_{idle,eap} = (1 - \tau_7)^{n_7} = 1 - P_{tran,eap} \\ P_{idle,rap} = \displaystyle\prod_{i=0}^{7}(1 - \tau_i)^{n_i} = 1 - P_{tran,rap} \end{cases} \quad (22)$$

The idle probability of a CSMA slot in a phase sensed by a $UP_i$ node is that the medium remains idle in a CSMA slot during the backoff procedure of the $UP_i$ node. Thus we have,

$$\begin{cases} P_{i,idle,eap} = \dfrac{(1 - \tau_i)^{n_i}}{(1 - \tau_i)} & ; \quad i = 7 \\ P_{i,idle,rap} = \dfrac{\prod_{i=0}^{7}(1 - \tau_i)^{n_i}}{(1 - \tau_i)} & ; \quad 0 \leq i \leq 7 \end{cases} \quad (23)$$

Now, we define $P_{i,idle}$ as the probability that a CSMA slot is idle sensed by a $UP_i$ node and the node decrements its backoff counter value by one considering counter lock probabilities.

$$\begin{cases} P_{i,idle} = P_{i,idle,rap}(1 - P_{i,lock}) & ; \quad 0 \leq i \leq 6 \\ P_{i,idle} = \left(\dfrac{L_{eap}}{L_{eap} + L_{rap}}P_{i,idle,eap} + \dfrac{L_{rap}}{L_{eap} + L_{rap}}P_{i,idle,rap}\right)(1 - P_{i,lock}) & ; \quad i = 7 \end{cases} \quad (24)$$

There is the significant difference between channel access probability and successful transmission probability. The channel access probability of $UP_i$ node in an access phase is the probability that the channel is successfully accessed by the $UP_i$ node during the access phase, conditioned on the fact that other nodes are not transmitting. Thus, we have

$$\begin{cases} P_{i,acce,eap} = \dfrac{n_i \tau_i (1 - \tau_i)^{n_i}}{(1 - \tau_i) P_{tran,eap}} & ; \quad i = 7 \end{cases} \quad (25)$$

34

$$P_{i,acce,rap} = \frac{n_i \tau_i \prod_{i=0}^{7}(1-\tau_i)^{n_i}}{(1-\tau_i)P_{tran,eap}} \quad ; \quad 0 \leq i \leq 7$$

Finally the channel access probability of $UP_i$ node is defined as

$$\begin{cases} P_{i,acce} = P_{i,acce,rap} & ; \quad 0 \leq i \leq 6 \\ P_{i,acce} = \frac{L_{eap}}{L_{eap}+L_{rap}}P_{i,acce,eap} + \frac{L_{rap}}{L_{eap}+L_{rap}}P_{i,acce,rap} & ; \quad i = 7 \end{cases} \quad (26)$$

The successful transmission probability of $UP_i$ node in an access phase is the probability that a $UP_i$ node access the channel successfully and receive ACK from the coordinator during the access phase. Thus, we have

$$\begin{cases} P_{i,succ,rap} = P_{i,acce,rap}(1-PER) & ; \quad 0 \leq i \leq 7 \\ P_{i,succ,eap} = P_{i,acce,eap}(1-PER) & ; \quad i = 7 \end{cases} \quad (27)$$

Finally, the successful transmission probability of $UP_i$ node is defined as

$$\begin{cases} P_{i,succ} = P_{i,succ,rap} & ; \quad 0 \leq i \leq 6 \\ P_{i,succ} = \frac{L_{eap}}{L_{eap}+L_{rap}}P_{i,succ,eap} + \frac{L_{rap}}{L_{eap}+L_{rap}}P_{i,succ,rap} & ; \quad i = 7 \end{cases} \quad (28)$$

Where $PER$ indicates the packet error rate whose value depends on access mechanism. We assume that immediate acknowledgement (I-ACK) is used in the network and a node transmits just one data frame during channel access period and thus, the packet error rate is written as

$$\begin{cases} PER = 1-(1-BER)^{L_{DATA}+L_{ACK}} & ; \quad basic \\ PER = 1-(1-BER)^{L_{RTS}+L_{CTS}+L_{DATA}+L_{ACK}} & ; \quad RTS/CTS \end{cases} \quad (29)$$

Where $BER$ denotes the bit error rate of the channel and $L_{RTS}$, $L_{CTS}$, $L_{DATA}$ and $L_{ACK}$ are the lengths of RTS, CTS, DATA and ACK frame in bits. After successfully access the channel, the transmission of a node is failed due to two reasons. The first reason is collision in transmission and the second one is error in transmission due to noisy channel. Therefore, the transmission failure probability of $UP_i$ node is written as

$$P_{i,fail} = P_{i,coll} + P_{i,error} \quad (30)$$

Where $P_{i,coll}$ and $P_{i,error}$ represent the transmission collision probability and transmission error probability, respectively. The transmission collision occurs when at least two nodes of an access phase transmit frame at the same time over the channel. The transmission collision probability of $UP_i$ node in an access phase is written as

$$\begin{cases} P_{i,coll,rap} = 1 - \frac{n_i \tau_i \prod_{i=0}^{7}(1-\tau_i)^{n_i}}{(1-\tau_i)P_{tran,rap}} & ; \quad 0 \leq i \leq 7 \\ P_{i,coll,eap} = 1 - \frac{n_i \tau_i (1-\tau_i)^{n_i}}{(1-\tau_i)P_{tran,eap}} & ; \quad i = 7 \end{cases} \quad (31)$$

Finally, the transmission collision probability of $UP_i$ node is defined as

$$\begin{cases} P_{i,coll} = P_{i,coll,rap} & ; \quad 0 \leq i \leq 6 \\ P_{i,coll} = \frac{L_{eap}}{L_{eap}+L_{rap}}P_{i,coll,eap} + \frac{L_{rap}}{L_{eap}+L_{rap}}P_{i,coll,rap} & ; \quad i = 7 \end{cases} \quad (32)$$

The transmission error occurs when the channel is accessed successfully as well as the frame is transmitted but the frame contains erroneous bit. The transmission error probability is written as

$$\begin{cases} P_{i,error,rap} = P_{i,acce,rap}PER & ; \quad 0 \leq i \leq 7 \\ P_{i,error,eap} = P_{i,acce,rap}PER & ; \quad i = 7 \end{cases} \quad (33)$$

Finally, the transmission error probability of $UP_i$ node is defined as

$$\begin{cases} P_{i,error} = P_{i,error,rap} & ; \ 0 \leq i \leq 6 \\ P_{i,error} = \dfrac{L_{eap}}{L_{eap} + L_{rap}} P_{i,error,eap} + \dfrac{L_{rap}}{L_{eap} + L_{rap}} P_{i,error,rap} & ; \ i = 7 \end{cases} \quad (34)$$

By substituting Eq. (32) and Eq. (34) in Eq. (30), the failure transmission probability can be obtained as

$$\begin{cases} P_{i,fail} = 1 - P_{i,acce,rap}(1 - PER) & ; \ 0 \leq i \leq 6 \\ P_{i,fail} = \dfrac{L_{eap}}{L_{eap} + L_{rap}}(1 - P_{i,acce,eap}(1 - PER)) + \dfrac{L_{rap}}{L_{eap} + L_{rap}} & ; \ i = 7 \\ \qquad (1 - P_{i,acce,rap}(1 - PER)) \end{cases} \quad (35)$$

In this work, since we assumed that packet arrival rate of $UP_i$ node is Poisson process with rate $\lambda_i$, the probability that node has at least one packet waiting for transmission can be determined as

$$\rho_i = 1 - e^{-\lambda_i T_{e,i}} \quad (36)$$

Where $T_{e,i}$ represents the expected time spent by the $UP_i$ node at each state. The duration of this time is not fixed and depends on the channel state, transmission state and access phase. If the channel is idle, the duration of the state is one CSMA slot, $T_{csmaslot}$. When the channel is sensed as busy its means that either successful transmission, transmission collision, or error transmission is occurred in the channel. If successful transmission is occurred, the duration of the state is the time of a successful transmission, which is estimated as

$$\begin{cases} T_{e,i} = T_{e,rap} & ; \ 0 \leq i \leq 6 \\ T_{e,i} = \dfrac{L_{eap}}{L_{eap} + L_{rap}} T_{e,eap} + \dfrac{L_{rap}}{L_{eap} + L_{rap}} T_{e,rap} & ; \ i = 7 \end{cases} \quad (37)$$

Where $T_{e,eap}$ and $T_{e,rap}$ represent the expected time spent by a node at each state during EAP and RAP phase, respectively.

$$T_{e,eap} = T_{csmaslot}(1 - P_{tran,eap}) + T_{succ} P_{tran,rap} P_{7,succ,eap} \\ + T_{coll} P_{tran,rap} P_{7,coll,eap} + T_{error} P_{tran,rap} P_{7,error,eap} \quad (38)$$

$$T_{e,rap} = T_{csmaslot}(1 - P_{tran,rap}) + T_{succ} P_{tran,rap} \sum_{i=0}^{7} P_{i,succ,rap} \\ + T_{coll} P_{tran,rap} \sum_{i=0}^{7} P_{i,coll,rap} + T_{error} P_{tran,rap} \sum_{i=0}^{7} P_{i,error,rap} \quad (39)$$

The duration of successful transmission ($T_{succ}$), collision transmission ($T_{coll}$) and error transmission ($T_{error}$) depends on access mechanism.

$$\begin{cases} T_{coll} = T_{DATA} + T_{ACK} + T_{SIFS} + 2\alpha & ; \ basic \\ T_{coll} = T_{RTS} + T_{CTS} + T_{SIFS} + 2\alpha & ; \ RTS/CTS \end{cases} \quad (40)$$

$$\begin{cases} T_{succ} = T_{error} = T_{DATA} + T_{ACK} + T_{SIFS} + 2\alpha & ; \ basic \\ T_{succ} = T_{error} = T_{RTS} + T_{CTS} + T_{DATA} + T_{ACK} + 3T_{SIFS} + 4\alpha & ; \ RTS/CTS \end{cases} \quad (41)$$

where $\alpha$ denotes the propagation delay and $T_{RTS}$, $T_{CTS}$, $T_{DATA}$, $T_{ACK}$ and $T_{SIFS}$ represent the duration of RTS, CTS, DATA, ACK and SIFS in time, respectively and that are determined as

$$T_{DATA} = \dfrac{Preamble}{R_S} + \dfrac{PHY\ Header}{R_{PLCP}} + \dfrac{(MAC\ Header + Framebody + FCS) \times 8}{R_{PSDU}} \quad (42)$$

$$T_{RTS/CTS/ACK} = \frac{Preamble}{R_S} + \frac{PHY\ Header}{R_{PLCP}} + \frac{(MAC\ Header + FCS) \times 8}{R_{PSDU}} \quad (43)$$

Simplifying all the above derived equations of the proposed analytical model, we obtain 43 equations while we have 43 unknown variables: $P_{idle,eap}$, $P_{idle,rap}$, $P_{i,idle}$, $P_{7,acce,eap}$, $P_{i,acce,rap}$, $P_{i,fail}$, $\tau_i$ and $\rho_i$. Finally, we solve the simplified equations using Maple [30] and then determine the performance metrics that are defined in following Section.

## 5. PERFORMANCE METRICS

### 5.1. Reliability

The reliability ($R_i$) of $UP_i$ node is defined as the complementary probability with which a transmitted packet is dropped due to repeated failure transmission after $m_i + x_i + 1$ attempts [10]. Therefore, $R_i$ is mathematically expressed as

$$R_i = 1 - P_{i,drop} = 1 - (P_{i,fail})^{m_i + x_i + 1} \quad (44)$$

### 5.2. Normalized Throughput

The normalized throughput ($S_i$) of $UP_i$ node is defined as the ratio of the time that the $UP_i$ node uses for successful transmission of data frame at a state and the time the $UP_i$ node stays at that state [18]. Therefore, $S_i$ is mathematically expressed as

$$\begin{cases} S_i = \dfrac{P_{i,succ} P_{tran,rap} X_{rap} L_{framebody}}{L_{eap} + L_{rap}} & ;\ 0 \le i \le 6 \\ S_i = \dfrac{P_{i,succ} (P_{tran,eap} X_{eap} + P_{tran,rap} X_{rap}) L_{framebody}}{L_{eap} + L_{rap}} & ;\ i = 7 \end{cases} \quad (45)$$

Where $L_{framebody}$ is the size of payload/frame body size in slots, $X_{rap}$ represents the average number of slots during RAP and $X_{eap}$ indicates the average number of slots during EAP. $X_{eap}$ and $X_{rap}$ are approximated as

$$X_{eap} = \frac{L_{eap}}{L_{e,eap}} \text{ and } X_{rap} = \frac{L_{rap}}{L_{e,rap}} \quad (46)$$

### 5.3. Energy Consumption

The energy consumption of a node is affected by several stages: 1) idle stage, 2) successful stage, 3) collision stage and 4) error stage [10],[13],[21],[22]. Assume that $P_{TX}$, $P_{RX}$ and $P_{IDLE}$ denote the energy consumption in transmitting state, receiving state and idle state of a node, respectively. The mean energy consumption ($E_i$) of $UP_i$ node is expressed as

$$E_i = E_{i,idle} + E_{i,succ} + E_{i,coll} + E_{i,error} \quad (47)$$

where $E_{i,idle}$, $E_{i,succ}$, $E_{i,coll}$ and $E_{i,error}$ are the energy consumption during idle stage, successful stage, collision stage, and error stage, respectively.

$$E_{i,idle} = T_{csmaslot} P_{IDLE} P_{i,idle} \quad (48)$$

The Equation (49) and (50) present the energy consumption during different stages for basic and RTS/CTS access mechanisms, respectively.

$$\{\quad E_{i,succ} = (T_{DATA} P_{TX} + T_{ACK} P_{RX} + T_{SIFS} P_{IDLE}) P_{tran} P_{i,succ} \quad (49)$$

$$+T_{succ}P_{IDLE}P_{tran}\left(\sum_{i=0}^{7} P_{i,succ} - P_{i,succ}\right)$$
$$E_{i,coll} = (T_{DATA}P_{TX} + T_{ACK}P_{RX} + T_{SIFS}P_{IDLE})P_{tran}P_{i,coll}$$
$$+T_{coll}P_{IDLE}P_{tran}\left(\sum_{i=0}^{7} P_{i,coll} - P_{i,coll}\right)$$
$$E_{i,error} = (T_{DATA}P_{TX} + T_{ACK}P_{RX} + T_{SIFS}P_{IDLE})P_{tran}P_{i,error}$$
$$+T_{succ}P_{IDLE}P_{tran}\left(\sum_{i=0}^{7} P_{i,error} - P_{i,error}\right)$$

$$\begin{cases} E_{i,succ} = (T_{RTS}P_{TX} + T_{CTS}P_{RX} + T_{DATA}P_{TX} + T_{ACK}P_{RX} \\ \qquad\qquad + 3T_{SIFS}P_{IDLE})P_{tran}P_{i,succ} \\ \qquad +T_{succ}P_{IDLE}P_{tran}\left(\sum_{i=0}^{7} P_{i,succ} - P_{i,succ}\right) \\ E_{i,coll} = (T_{RTS}P_{TX} + T_{CTS}P_{RX} + T_{SIFS}P_{IDLE})P_{tran}P_{i,coll} \\ \qquad +T_{coll}P_{IDLE}P_{tran}\left(\sum_{i=0}^{7} P_{i,coll} - P_{i,coll}\right) \\ E_{i,error} = (T_{RTS}P_{TX} + T_{CTS}P_{RX} + T_{DATA}P_{TX} + T_{ACK}P_{RX} \\ \qquad\qquad + 3T_{SIFS}P_{IDLE})P_{tran}P_{i,error} \\ \qquad +T_{succ}P_{IDLE}P_{tran}\left(\sum_{i=0}^{7} P_{i,error} - P_{i,error}\right) \end{cases} \quad (50)$$

### 5.4. Average Access Delay

The average access delay of $UP_i$ node is defined as the average time from the instant when frame is generated to the instant when the packet is successfully transmitted or dropped. The average access delay has two parts 1) waiting time for permitted access phase 2) channel access time to transmit frame [29]. The waiting time is calculated as follows:

$$T_{wait} = \frac{T_{eap}}{2} \quad (51)$$

The successful transmission is occurred at state $(i,j,0)$ if a collision occurs at each state $(i,l,0)$, $1 \leq l \leq j-1$, and no collision occurs at state $(i,j,0)$. Packet drop is occurred if a collision occurs at each state $(i, m_i + x_i, 0)$. Thus the average access delay of $UP_i$ node can be determined as

$$\begin{cases} T_{i,delay} = T_{wait} + \sum_{j=0}^{m_i+x_i}(P_{i,fail})^j(1-P_{i,fail})\sum_{l=0}^{j}\frac{W_{i,l}+1}{2}T_{e,i} \\ \qquad\qquad + (P_{i,fail})^{m_i+x_i+1}\sum_{l=0}^{m_i+x_i+1}\frac{W_{i,l}+1}{2}T_{e,i} \\ T_{i,delay} = \sum_{j=0}^{m_i+x_i}(P_{i,fail})^j(1-P_{i,fail})\sum_{l=0}^{j}\frac{W_{i,l}+1}{2}T_{e,i} \qquad ; \quad i=7 \end{cases} \quad ;\ 0 \leq i \leq 6 \quad (52)$$

$$+\left(P_{i,fail}\right)^{m_i+x_i+1} \sum_{l=0}^{m_i+x_i+1} \frac{W_{i,l}+1}{2} T_{e,i}$$

## 6. PERFORMANCE EVALUATION

We consider a general experimental scenario of intra-WBAN to investigate the effects of packet arrival rate, channel condition, payload size, access phase length, access mechanism and number of nodes on performance metrics of IEEE 802.15.6 CSMA/CA MAC protocol. We assume that the length of MAP1, EAP2, RAP2, MAP2 and CAP of the superframe are set to zero. The general parameters of the intra-WBAN are given in Table 4.

Table 4. MAC and PHY attributes of the experiment

| Parameters | Values | Parameters | Values |
|---|---|---|---|
| Preamble | 90 bits | Data rate for PLCP | 91.9 kbps |
| PHY Header | 31 bits | Data rate for PSDU | 971.4 kbps |
| MAC Header | 56 bits | RTS | 193 bits |
| FCS | 16 bits | ACK | 193 bits |
| Frequency band | 2.4 GHz | CTS | 193 bits |
| Retry limit | 7 | $P_{TX}$ | 27 mW |
| Symbol rate | 600 ksps | $P_{RX}$ | 1.8 mW |
| Propagation delay | 1 µs | $P_{IDLE}$ | 5 µW |
| SIFS | 75 µs | CSMA slot | 125 µs |

### 6.1. Effects of Arrival Rate

In order to investigate the effects of arrival rate on the performance metrics, we consider a network scenario that operates with RTS/CTS access mechanism, non-saturation traffic condition of nodes and noisy channel ($BER = 2 \times 10^{-5}$). The network scenario consists of 16 nodes where each user priorities contain 2 nodes. We set the payload size as $Framebody = 100\ bytes$ and access phase lengths as $eap1 = 0.1\ sec$ and $rap1 = 0.8\ sec$ while other parameters are same as general scenario. We assume that all user priorities have same packet arrival rate, i.e., $\lambda_0 = \lambda_1 = \cdots = \lambda_7$. The effects of arrival rate vary from 0.5 $pkts/sec$ to 4.0 $pkts/sec$, on the performance metrics viz. reliability, normalized throughput, energy consumption and average access delay, are illustrated in Figure 5.

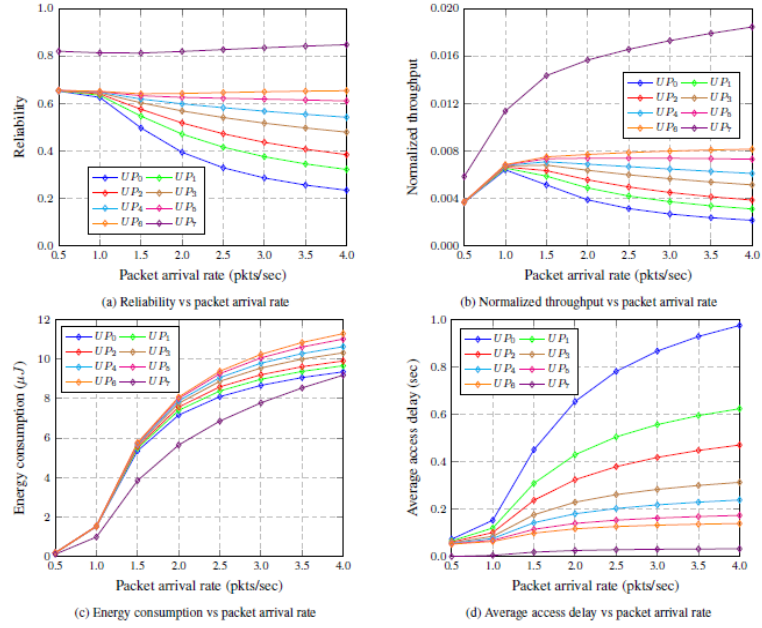

Figure 5.   Effects of packet arrival rate on the performance metrics

The results in Figure 5 (a) and (b) show that as the packet arrival rate increases, the reliability and normalized throughput of lower user priorities ($UP_0$ –$UP_6$) decrease whereas the reliability and normalized throughput of highest user priorities ($UP_7$) increases. This is because the collision probability increases with the packet arrival rate increasing. It is also observed that energy consumption and average access delay exponentially increase with the packet arrival rate increasing in Figure 5 (c) and (d). This is due to the fact that the collision probability increases as the packet arrival rate increases.

## 6.2. Effects of Channel Condition

In order to investigate the effects of channel conditions on the performance metrics, we consider a network scenario that operates with RTS/CTS access mechanism and saturation traffic condition. The channel condition is expressed by the value of bit error rate ($BER$) where $BER = 0$ represents the ideal channel and $BER > 0$ represents the noisy channel. The network scenario consists of 16 nodes where each user priorities contain 2 nodes. We set the payload size as $Framebody = 100\ bytes$ and access phase length as $eap1 = 0.1\ sec$ and $rap1 = 0.8\ sec$ while other parameters are same as general scenario. The effects of channel condition (BER), varies from $0 - 0.01$ , on the performance metrics viz. reliability, normalized throughput, energy consumption and average access delay, are illustrated in Figure 6.

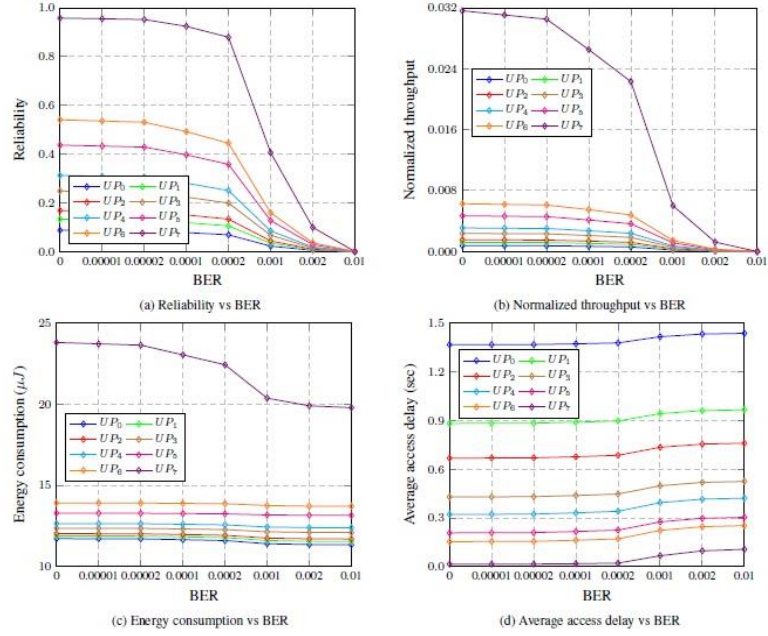

Figure 6. Effects of channel condition on the performance metrics

The results show that the reliability, normalized throughput and energy consumption of all UPs are maximum whereas the average access delay is minimum for the ideal ($BER = 0$) channel condition. This is because the collision probability is almost constant for fixed number of nodes in the network and there is no failure transmission occurred due to transmission error. With the bit error rate increasing, it would cause many error transmissions, which resulting in a significant decrease in the reliability, normalized throughput and energy consumption as well as a slight increase in the average access delay. It is also observed that the performance metrics is almost stable up to $BER = 2 \times 10^{-4}$ whereas with the bit error rate further increasing, the performance drastically degrades.

### 6.3. Effects of Payload Size

In order to investigate the effects of payload ($Framebody$) size on the performance metrics, we consider a network scenario that operates with RTS/CTS access mechanism and saturation traffic condition. The network scenario consists of 16 nodes where each user priorities contain 2 nodes. We set the bit error rate as $BER = 2 \times 10^{-5}$ and access phase lengths as $eap1 = 0.1\ sec$ and $rap1 = 0.8\ sec$ while other parameters are same as general scenario. The effects of payload size, varies from $0 - 260\ bytes$, on the performance metrics viz. reliability, normalized throughput, energy consumption and average access delay, are illustrated in Figure 7. The results in Figure 7 (a) and (d) show that payload size has no significant impact on reliability and access delay, respectively. This is because the collision probability is almost constant for fixed number of nodes in the network. It is also observed that the normalized throughput and energy consumption sharply increases as the frame payload size increases and highest user priority increases faster than other user priorities in Figure 7 (b) and (c). This is due to the fact that number of successfully transmitted bits increases as the payload size increases where the collision probability is almost constant and thus increase of energy consumption. The highest user nodes achieve more throughput and consume more power than others because $UP_7$ has transmission opportunities in both access phases (EAP1 and RAP1) with the lowest contention window bound.

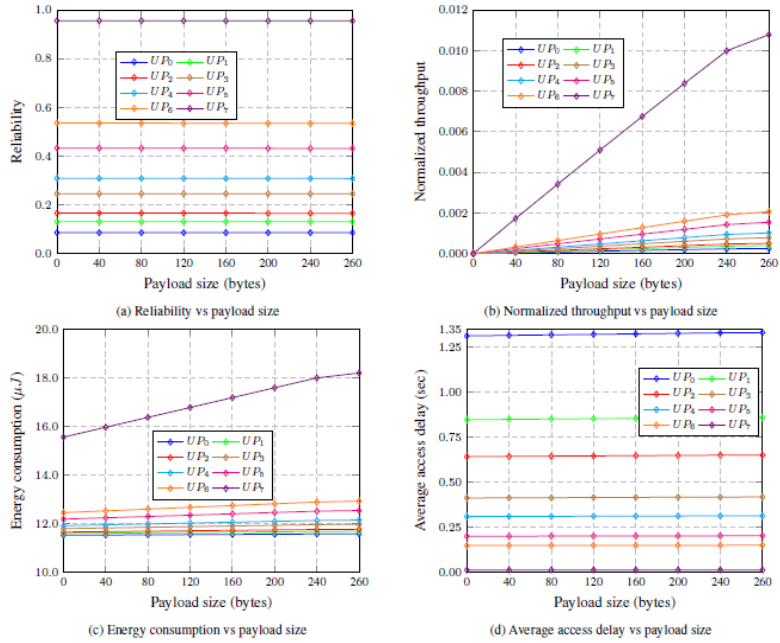

Figure 7. Effects of payload size on the performance metrics

## 6.4. Effects of Access Phase Length

In order to investigate the effects of access phase length on the performance metrics, we consider a network scenario that operates with RTS/CTS access mechanism, non-saturation traffic condition and noisy channel ($BER = 2 \times 10^{-5}$). The network scenario consists of 16 nodes where each user priorities contain 2 nodes. We set the payload size as $Framebody = 100\ bytes$ and exclusive access phase length as $eap1 = 0.1\ sec$ while other parameters are same as general scenario. We assume that the packet arrival rate of each user priorities is equal, i.e, $\lambda_0 = \lambda_1 = \cdots = \lambda_7 = 2.0\ pkts/sec$. We vary the length of RAP1 from $0.1\ sec - 0.8\ sec$ because all user priorities nodes can access the channel during the random access phase (RAP) only. The effects of access phase length (RAP1) on the performance metrics viz. reliability, normalized throughput, energy consumption and average access delay, are illustrated in Figure 8. The results in Figure 8 (b) and (c) show that equal length of EAP1 and RAP1 (eap1=rap1=0.1 sec) gives an aggressive prioritization in $UP_7$ over the other UPs. As the length of RAP1 increases, the difference in normalized throughput and energy consumption between $UP_7$ and other lower user priorities decreases. It is also noted that the normalized throughput and energy consumption of $UP_7$ exponentially decrease whereas the normalized throughput and energy consumption of other user priorities gradually increase. With the length of RAP1 increasing, the reliability gradually decreases while the average access delay gradually increases as shown in Figure8 (a) and (d).

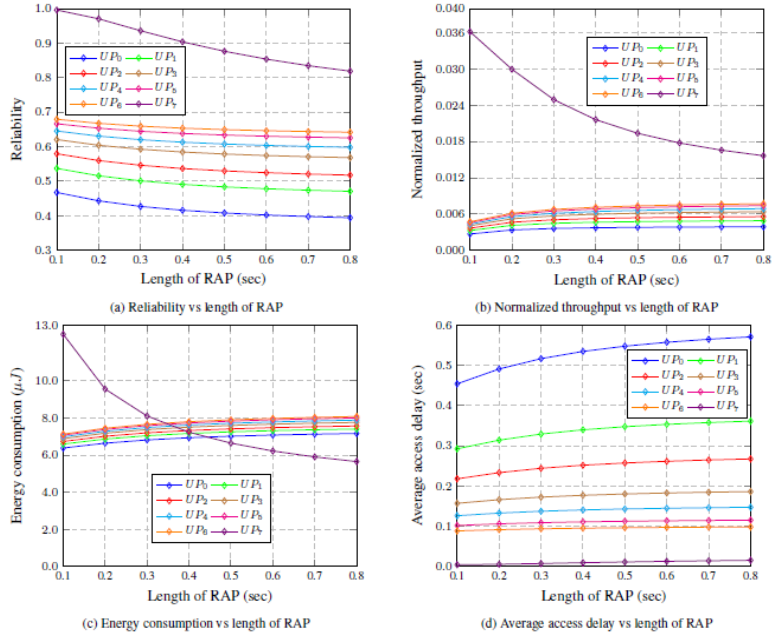

Figure 8. Effects of access phase length on the performance metrics

## 6.5. Effects of Access Mechanism and Number of Nodes

In order to investigate the effects of access mechanism and number of nodes on the performance metrics, we consider a network scenario that operates with non-saturation traffic condition and noisy channel ($BER = 2 \times 10^{-5}$). We set the payload size as $Framebody = 100\ bytes$ and access phase lengths as $eap1 = 0.1\ sec$ and $rap1 = 0.8\ sec$ while other parameters are same as general scenario. We also assume that all user priorities have same packet arrival rate, i.e., $\lambda_0 = \lambda_1 = \cdots = \lambda_7 = 0.5\ pkts/sec$. The effects of access mechanism (basic and RTS/CTS) and number of nodes, varies from $8 - 64$, on the performance metrics viz. reliability, normalized throughput, energy consumption and average access delay, are illustrated in Figure 9 – 12. The results show that the reliability, normalized throughput, energy consumption and average access delay of the lower user priorities ($UP_0$ - $UP_6$) nodes are almost the same up to 24 and 32 nodes for the basic access mechanism and RTS/CTS access mechanism, respectively. It is also observed that as the number of nodes further increases, the reliability and normalized throughput sharply decrease whereas energy consumption and average access delay exponentially increase. This increasing rate or decreasing rate of performance metrics of basic access mechanism is more than the RTS/CTS access mechanism. This is because the collision probability increases slowly with the number of nodes increasing within a certain range (24 for the basic access mechanism and 32 for the RTS/CTS access mechanism) whereas with the number of vehicles further increasing, the collision probability increases dramatically. Moreover, the increasing rate of collision probability of the basic access mechanism is more than the RTS/CTS access mechanism. The results also show that the highest user priority ($UP_7$) nodes gain the highest reliability (almost 80%), achieve the highest normalized throughput (on average 0.8%), consume the lowest energy (on average 5 µJ) and enjoy the lowest (almost zero) average access delay compared to the lower user priorities ($UP_0$ - $UP_6$) nodes. This is because $UP_7$ nodes are assigned an aggressive prioritization over the other UPs by providing transmission opportunities in both access phases (EAP1 and RAP1) with the lowest contention window bound. It is also shown that the RTS/CTS access mechanism improves the performance of $UP_7$ nodes in all aspects compared to the basic access mechanism.

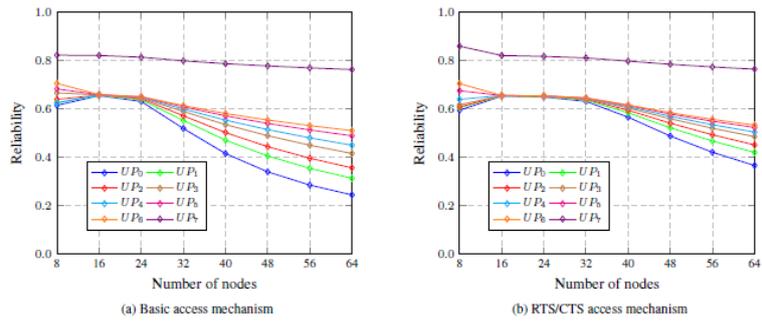

Figure 9. Effects of access mechanisms and number of nodes on the reliability

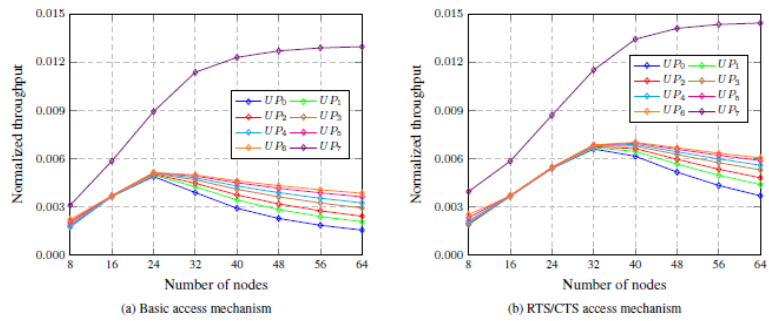

Figure 10. Effects of access mechanism and number of nodes on the normalized throughput

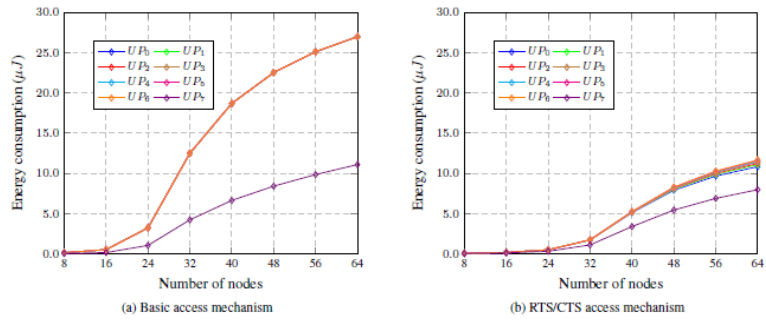

Figure 11. Effects of access mechanisms and number of nodes on the energy consumption

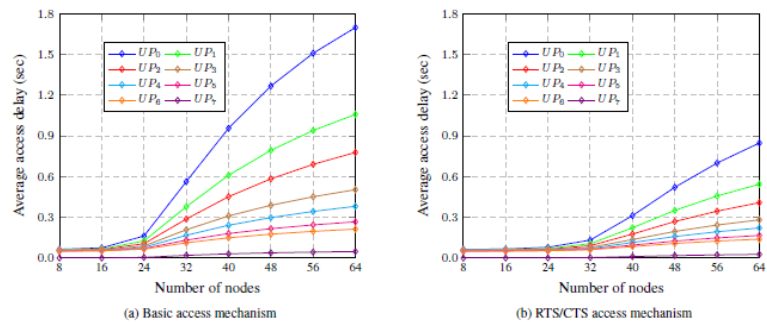

Figure 12. Effects of access mechanism and number of nodes on the average access delay

## 7. CONCLUSION

In this paper, we developed a Markov chain-based analytical model of IEEE 802.15.6 CSMA/CA and solved the complex analytical model by Maple to investigate the effects of the packet arrival rate, channel condition, payload size, access phase length, access mechanism and number of nodes on the performance parameters viz. reliability, normalized throughput, energy consumption and average access delay. The analytical results clearly indicated that different MAC and PHY parameters have significant impact on the performance of IEEE 802.15.6 CSMA/CA. Finally, we concluded the necessity of different access phases, access mechanisms and user priorities in IEEE 802.15.6 standard. We believe that the evaluated results would assist the communication protocol developers to design different algorithms and to select optimal value of different MAC and PHY parameters based on desired QoS.